**Preseismic oscillating electric field "strange attractor like" precursor, of T = 6 months, triggered by Ssa tidal wave. Application on large (Ms > 6.0R) EQs in Greece (October 1st, 2006 - December 2nd, 2008).**

Thanassoulas[1], C., Klentos[2], V., Verveniotis, G.[3], Zymaris, N.[4]

1. Retired from the Institute for Geology and Mineral Exploration (IGME), Geophysical Department, Athens, Greece.
   e-mail: thandin@otenet.gr - URL: www.earthquakeprediction.gr

2. Athens Water Supply & Sewerage Company (EYDAP),
   e-mail: klenvas@mycosmos.gr - URL: www.earthquakeprediction.gr

3. Ass. Director, Physics Teacher at 2nd Senior High School of Pyrgos, Greece.
   e-mail: gver36@otenet.gr - URL: www.earthquakeprediction.gr

4. Retired, Electronic Engineer.

**Abstract.**

In this work the preseismic "strange attractor like" precursor is studied, in the domain of the Earth's oscillating electric field for **T = 6 months**. It is assumed that the specific oscillating electric field is generated by the corresponding lithospheric oscillation, triggered by the **Ssa** tidal wave of the same wave length (**6 months**) under excess strain load conditions met in the focal area of a future large earthquake. The analysis of the recorded Earth's oscillating electric field by the two distant monitoring sites of **PYR** and **HIO** and for a period of time of 26 months (October 1st, 2006 - December 2nd, 2008) suggests that the specific precursor can successfully resolve the predictive time window in terms of months and for a "swarm" of large **EQs (Ms > 6.0R),** in contrast to the resolution obtained by the use of electric fields of shorter (**T = 1, 14 days**, single **EQ** identification) wave length. More over, the fractal character of the "strange attractor like" precursor in the frequency domain is pointed out. Finally, a proposal is made that concerns the continuous monitoring of the specific preseismic attractor in distinct different wave lengths of the oscillating Earth's electric field so that an early warning system can be utilized. As a refinement of the "strange attractor like" methodology, the guide lines of a generalized inversion scheme are presented so that the epicenter of the driving mechanism (seismic epicentral area) can be estimated in a least squares sense.

**1. Introduction.**

The role of the lithospheric tidal waves in the triggering of the earthquakes has been recognized long ago. Thanassoulas (2007) presented a review of the literature on this topic along with examples of the detailed holding mechanism too. Furthermore, it has been shown that, the very same tidal waves, when combined to the large scale piezoelectric mechanism triggered in the lithosphere, it is capable to produce various types of electric signals (Thanassoulas 2007, Thanassoulas 2008). The specific case of the Earth's oscillating electric field can be used for the compilation of "phase maps" (Nusse and Yorke, 1998; Korsch, Jodl, and Hartmann, 2008) from its components that are registered at distant monitoring sites. Phase maps were compiled by the use of this method for the cases of the Earth's oscillating fields registered at **PYR** and **HIO** monitoring sites in Greece with periods of **T = 1 day** and **T = 14 days**. The latter correspond to the **K1, P1** and **M1** tidal components. The study of these maps suggested that a quite short time predictive window, of a few days, can be determined before the occurrence of large (**Ms > 6.0R**) **EQs**. The latter could be utilized by the detection of the "strange attractor like" preseismic precursor (Thanassoulas et al. 2008a, b, c).

Moreover, it was shown (Thanassoulas, 2007) that the total seismicity of a large seismogenic area varies following closely the longer tidal wavelengths, such as of **T = 6 or 12 months,** of the oscillating lithosphere. Therefore, "strange attractor like" preseismic precursors should be, hopefully, detected before large **EQs** corresponding to the same long wavelength of the Earth's oscillating electric field. To this end an experimental initial analysis was performed of the Earth's oscillating electric field, at **T = 6 months,** which corresponds to the **Ssa** tidal component, aiming into detection of long period "strange attractor like" preseismic precursors. In this case too, the registrations of the Earth's electric field by **PYR** and **HIO** monitoring sites were used.

**2. The data.**

The study period spans from October 1st, 2006 to December 4th, 2008. The location of the **PYR** and **HIO** monitoring sites is presented in figure **(1)**.

During this time period eight **(8)** large earthquakes occurred with magnitude **Ms > 6.0R**. Their date, depth of occurrence, and magnitude are tabulated in **TABLE – 1.** These EQs are numbered from **(1)** to **(8).** Each number denotes the corresponding **EQ** in the map of figure **(1).**

**TABLE – 1.**

| Earthquake | Depth (in Km) | Magnitude (Ms) |
|---|---|---|
| 1. March 25$^{th}$, 2007 | 15 | 6.0 |
| 2. January 6$^{th}$, 2008 | 86 | 6.6 |
| 3. February 14$^{th}$, 2008 | 41 | 6.7 |
| 4. February 20$^{th}$, 2008 | 25 | 6.5 |
| 5. June 8$^{th}$, 2008 | 25 | 7.0 |
| 6. June 21$^{st}$, 2008 | 12 | 6.0 |
| 7. July 15$^{th}$, 2008 | 56 | 6.7 |
| 8. October 14$^{th}$, 2008 | 24 | 6.1 |

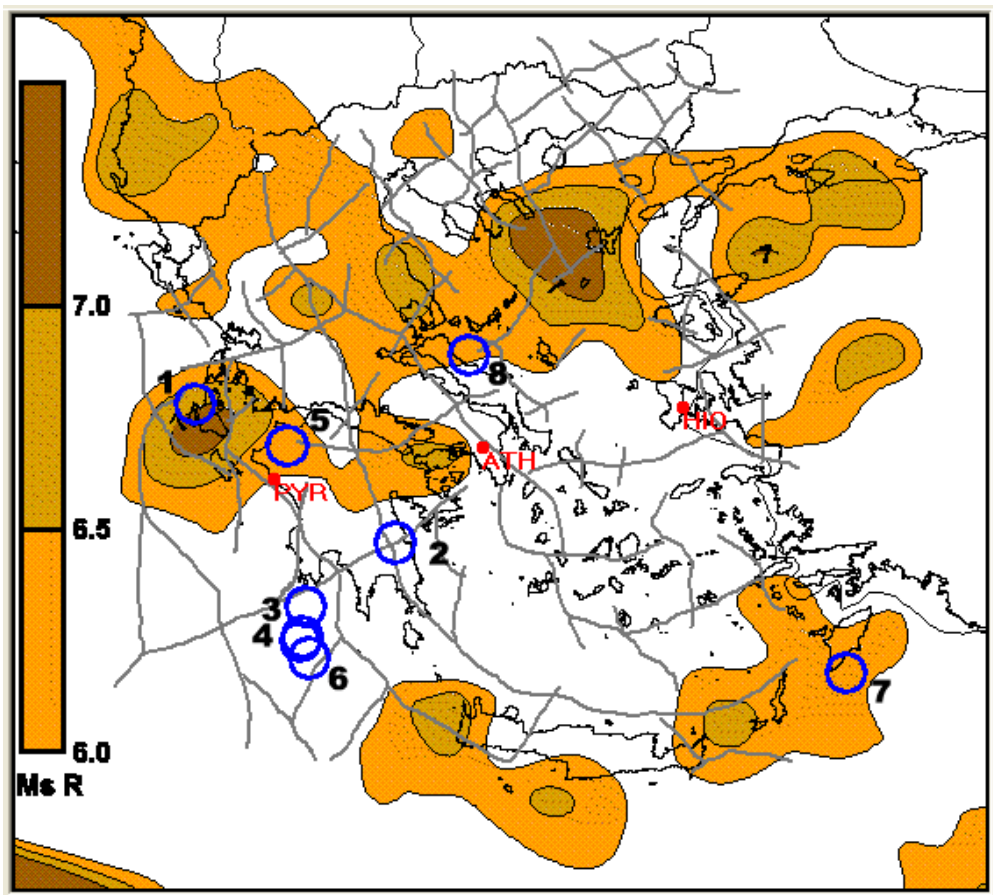

Fig. 1. Location (blue circles) of the EQs of **Ms > 6.0R** which occurred in Greece during October 1$^{st}$, 2006 to December 4$^{th}$, 2008. The numbers refer to **TABLE - 1**. The colour coded areas correspond to areas of large seismic potential (expressed in equivalent **EQ** magnitude) while the thick grey lines depict the deep lithospheric fracture zones calculated from the inversion of the Earth's gravity field (Thanassoulas, 2007).

The Earth's electric field, which was recorded during the study period by the **PYR** monitoring site, is presented in figure **(2).** The large **EQs** which occurred in the same period of time are denoted by red bars.

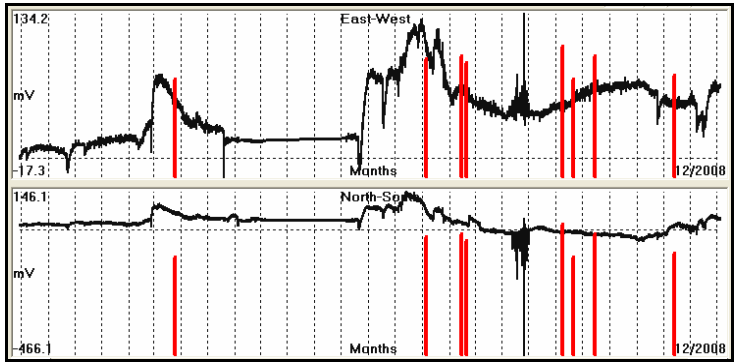

Fig. 2. Raw data (black line) of the Earth's electric field recorded by **PYR** monitoring site during the period from October 1[st], 2006 to December 4[th], 2008. The red bars indicate the occurrence of large (**Ms > 6.0R**) **EQs**.

The data of figure **(1)** were band-pass filtered **(FFT)** with a center filter period of **Tc = 6 months**, a bandwidth of **12** months and are presented in figure (**2a**).

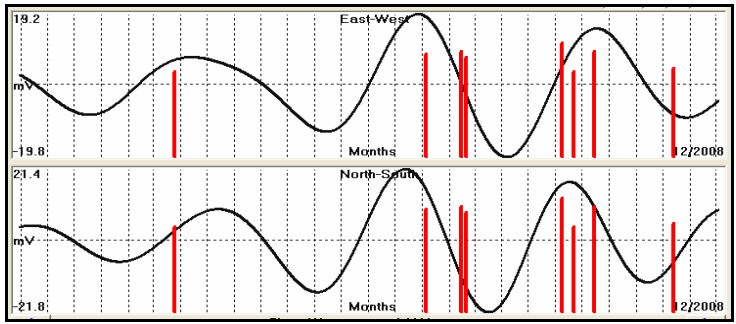

Fig. 2a. Oscillating data (**Tc = 6 months**, black line) of the Earth's electric field obtained from **PYR** raw data (after band-pass filtering) during the period from October 1[st], 2006 to December 4[th], 2008. The red bars indicate the occurrence of large (**Ms > 6.0R**) **EQs**.

The Earth's electric field which was recorded, during the study period, by the **HIO** monitoring site is presented in figure **(3)**. The large **EQs** that occurred in the same period are denoted by red bars.

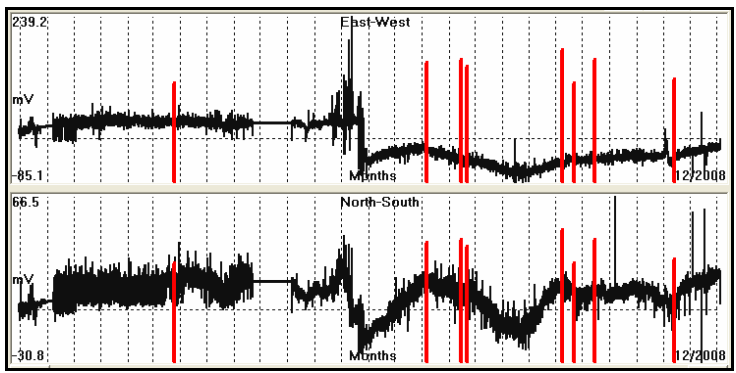

Fig. 3. Raw data (black line) of the Earth's electric field recorded by **HIO** monitoring site during the period from October 1[st], 2006 to December 4[th], 2008. The red bars indicate the occurrence of large (**Ms > 6.0R**) **EQs**.

The data of figure **(3)** were band-pass filtered **(FFT)** with a center filter period of **Tc = 6 months**, a bandwidth of **12** months and are presented in figure (**3a**).

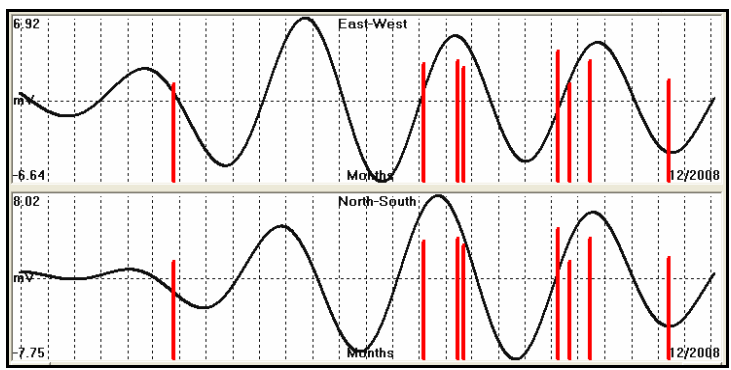

Fig. 3a. Oscillating data (**Tc = 6 months**, black line) of the Earth's electric field obtained from **HIO** raw data (after band-pass filtering) during the period from October 1[st], 2006 to December 4[th], 2008. The red bars indicate the occurrence of large (**Ms > 6.0R**) **EQs**.

## 3. Phase maps presentation.

The depicted data in figures (**2a**) and (**3a**) were used to compile the corresponding phase maps (Thanassoulas et al. 2008a) for time intervals related to the occurred **EQs**. For this purpose the original data (**1 sample per minute**) were resampled at a rate of one (**1)** sample per ten (**10**) minutes. This modification of the data was adopted so that the compilation of the maps would be drastically accelerated without loosing the details of the map. The compiled phase maps are presented in figures (**5, 6, 7**). The study period has been divided into three (**3**) sub-periods (**A, B, C**) according to whether a "strange attractor like" seismic precursor is present or not (**fig. 4**). The first period (**A**) spans from October 1st 2006 to October 1st 2007. The second sub-period (**B**) spans from October 1st 2007 – September 4th 2008, and the third one (**C**) spans from September 4th 2008 – December 4th 2008.

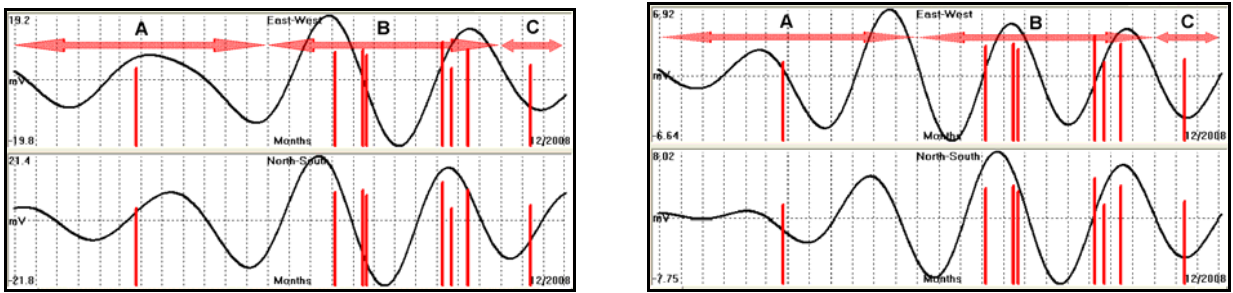

Fig. 4. Earth's oscillating electric field, of **T = 6 months,** determined from **PYR** (left) and **HIO** (right) monitoring sites. The double horizontal red arrows **(A, B, C)** denote the time sub-periods of absence or presence of the "strange attractor like" precursor.

The compiled phase maps are presented as follows:

### 3.1. October 1st 2006 – October 1st 2007 (time period A).

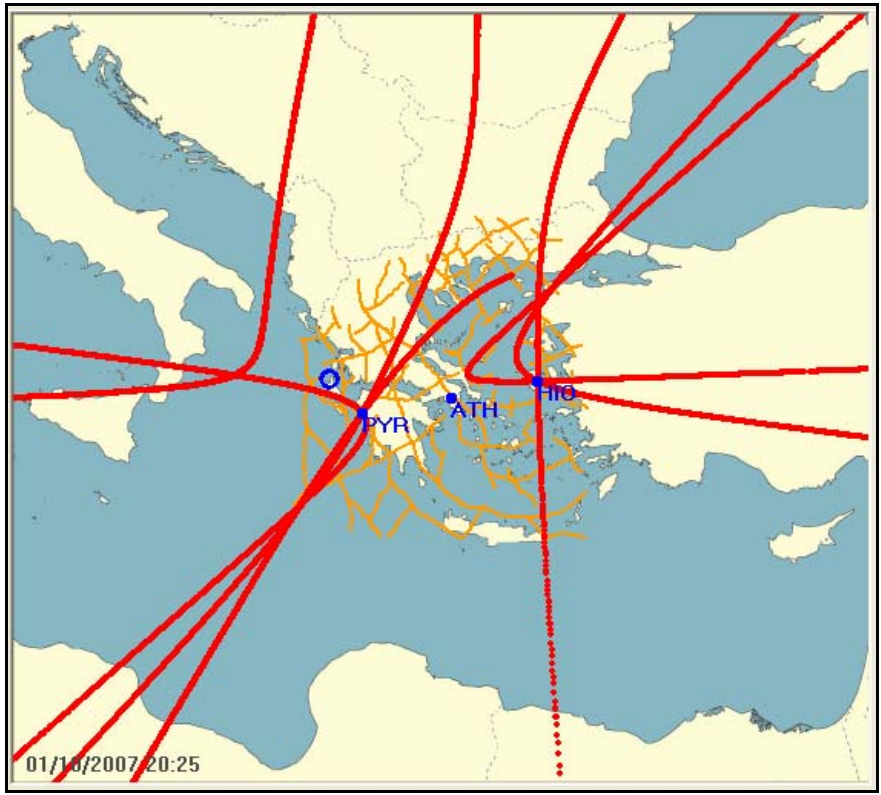

Fig. 5. Phase map calculated for the period of time from October 1st 2006 – October 1st 2007 (time period **A**). The blue circle denotes the **EQ** of March 25th 2007, **Ms = 6.0R**.

During this period of time (**12 months**) the phase map is characterized by the presence of hyperbolas only, while just one large seismic event (March 25th 2007, **Ms = 6.0R** ) took place.

## 3.2 October 1st 2007 – September 4th 2008 (time period B).

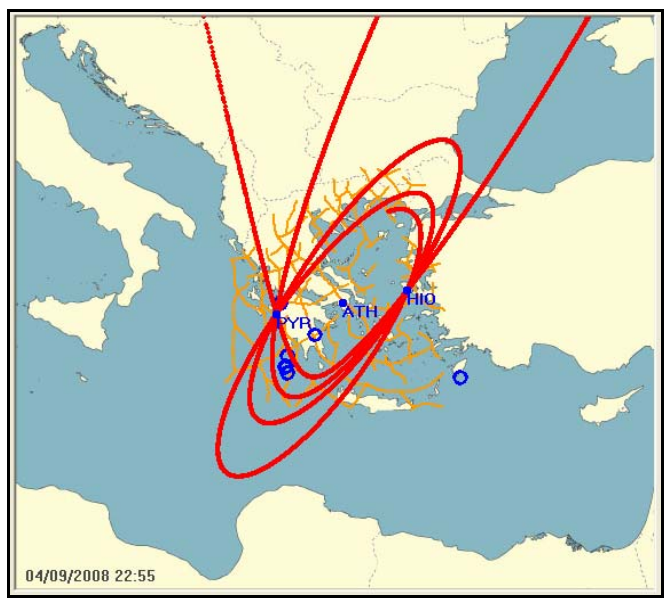

Fig. 6. Phase map calculated for the period of time from October 1**st** 2007 – September 4**th** 2008 (time period **B**). The blue circles denote the large (**Ms > 6.0R**) **EQs** (**2, 3, 4, 5, 6, 7 of TABLE-1**) which took place within this time period.

During this period of time (**11 months**) the phase map is characterized by the presence of ellipses only, while six (**6**) large seismic events (**2, 3, 4, 5, 6, 7 of TABLE-1**) took place. The occurrence of such a large number of strong **EQs** in to such short period of time is well out of the expected normal (about 1/year) seismicity of similar magnitudes of the regional seismogenic area of Greece.

## 3.3 September 4th 2008 – December 4th 2008 (time period C).

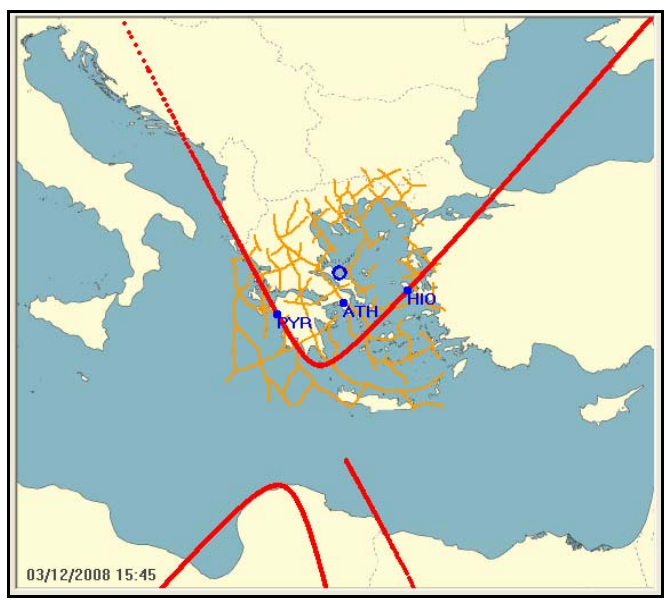

Fig. 7. Phase map calculated for the period of time from September 4**th** 2008 – December 4**th** 2008 (time period **C**). The blue circle denotes the **EQ** of October 14**th** 2008, **Ms = 6.1R**.
.
During this period of time (**3 months**) the phase map is characterized by the presence of hyperbolas only while just one large seismic event took place (October 14**th** 2008, **Ms = 6.1R**).

## 4. Discussions – conclusions.

In this study the **PYR** and **HIO** monitoring sites were used as in the cases of the Earth's oscillating field for **T = 1 day** and **T = 14 days** (Thanassoulas et al. 2008a, b, c). The latter was dictated by the fact that **PYR** and **HIO** receiving electric dipoles use very similar, in technical details, hardware (Thanassoulas, 2007). Consequently, the obtained results from the analysis of the Earth's oscillating electric field can be easily compared for the different **T** values.

In the present case and specifically for the sub-period of **A (fig. 5)**, it is clearly shown that no preseismic electric excitement of the regional seismogenic area is present at all. The phase map is characterized by hyperbolas for a period of **12 months** that corresponds to two (**2**) wavelengths of the used oscillating electric field. In this period of time only one **EQ** (March 25$^{th}$ 2007, **Ms = 6.0R**) took place. This seismic event fits the normal seismicity of Greece that is about one (**1**) such large event per **1 – 1.5** years. This single seismic event was not capable to trigger preseismic electric precursors of such large (**T = 6 months**) periods.

In the next sub-period **B (fig. 6)**, that follows immediately the one of **A**, the status of the regional seismogenic area changes drastically. For eleven (**11**) consecutive months, the Earth's oscillating electric field generates a "strange attractor like" preseismic precursor. The latter, following the theoretical part of the methodology (Thanassoulas 2007, Thanassoulas et al. 2008a) suggests the strong seismic excitation of the regional seismogenic area. Actually, during this sub-period, six (**6**) large **EQs** (**2, 3, 4, 5, 6, 7 of TABLE-1**) did occur. This number (**6**) of large **EQs** which did occur within such a short period of time is well above the expected normal yearly seismicity rate of large **EQs**. Therefore, it is quite logical to relate the presence of the ellipses ("strange attractor like" seismic precursors) during this period **(B)** with the large magnitude seismicity observed during the same period of time. Moreover, the latter is corroborated by the corresponding behaviour of sub-period **(A)**, when there were no "strange attractor like" precursors nor large seismicity too.

Finally, in the case of sub-period **C (fig. 7)**, all come back to normal. The seismic excitation of the regional seismogenic area, that concerns large seismic events, diminishes, the "strange attractor like" precursor vanishes too and only one **(1)** seismic event (October 14$^{th}$ 2008, **Ms = 6.1R**) takes place. Till the end of January 2009, when this work is typed, no any large **(M > 6.0R)** seismic event has taken place yet. Therefore, it is evident that the determined "strange attractor like" is closely related to the seismic excitement status of the under study period of time and to the **Ssa** tidal wave too.

The main difference between this study and the previous ones (Thanassoulas 2007, Thanassoulas et al. 2008a, b, c) is that the present study does not deal with the time of occurrence of a specific large **EQ** but instead deals rather with an extended period of time when the preseismic seismic stress load has increased over a wide regional seismogenic area. The latter has as an effect the generation of a sequence of large **EQs**. After reviewing the obtained results from the cases of the analysis of the Earth's oscillating field for **T = 1, 14 days** and **T = 6 months**, it is clear that the mechanism of the tidal waves and the oscillating lithosphere are the driving mechanisms for the generation of the "strange attractor like" precursor. The different tidal wave components trigger the generation of the "strange attractor like" precursor at the times when the stress load has reached critical values at a specific seismogenic area.

The triggering mechanism of the Earth's electric field generates a wide frequency spectrum of electric energy. Therefore, the "strange attractor like" properties should be normally present in all frequency bands when a seismogenic area has been preseismically electrically activated. This was verified by the presence of the "strange attractor like" precursor before certain large **EQs**, in both cases for **T = 1** and **14 days** simultaneously, while the same **EQs** were "identified" as a group of **EQs** by the "strange attractor like" that corresponds to the case of **T = 6 months**. The presence of the "strange attractor like" in almost all studied frequency bands fosters its validity as a seismic precursor.

The case of the "strange attractor like" precursor for **T = 6 months** corresponds to an integral behaviour of the wider regional seismogenic area that has been stress loaded in excess towards the generation of large **EQs**. Therefore, it can be considered as an electric precursor of a "swarm" of large **EQs** which will last for a rather long time period.

Viewing the specific strange attractor from the wavelength of the used oscillating electric field point of view, the achieved resolution that concerns the predictive time window for the occurrence of a large **EQ**, increases from long to short used wavelengths. The latter fits the general optics principles of physics that concerns the detectable details of an illuminated object by light of various wavelengths. The best resolution is obtained by using the Earth's oscillating electric field of **T = 1 day**. In this case the day of occurrence can be determined quite accurately (Thanassoulas et al. 2008b) in terms of a couple of days while if a longer wave length is used (**T = 14 days**) then the predictive time window becomes larger (always in terms of a few days, Thanassoulas et al. 2008c). In the case of **T = 6 months** the predictive window becomes so large that it is not possible to correspond it to any specific large **EQ**, but it rather can be related to the generation of a "swarm" like group of large **EQs** in a rather similar period of time. In this case the predictive scheme is no longer short-term but it becomes, more or less, a short to medium term one.

Another interesting feature of the "strange attractor like" precursor is its relation to the physical dimensions of the seismogenic area that are electrically preseismically activated. In the case of the very long wavelength (**T = 6 months**) oscillating electric field it is suggested that an integral large volume of the regional seismic area has been activated and therefore each individual large **EQ** and its surrounding neighbour seismogenic area behave as a very basic element absorbed in the whole process. On the other hand, for short wavelengths of the electric field (**T = 1 day**) each large **EQ** behaves as an individual integral case. Consequently, the "strange attractor like" precursor can be considered that behaves closely to fractal mode concerning the different wave lengths of the used oscillating electric field and the dimensions of the corresponding seismically activated lithospheric blocks, for each wavelength case.

As a result, of its fractal behaviour, the "strange attractor like" precursor can be used in various predictive modes, according to the wavelength of the used oscillating electric field, so that different predictive time windows (from short to medium term) can be achieved towards the earthquake prediction. In a real application of the methodology, the "strange attractor like" precursor must be continuously monitored in various modes of wavelengths of the recorded Earth's oscillating electric field. The early warning sign for a seismically dangerous excitement of a wide seismogenic area will firstly be identified by the longer wave length precursors. Following, shorter predictive windows for specific large **EQs** will be identified by the "strange attractor like" precursors which will be calculated from shorter wave lengths of the Earth's oscillating electric field.

At the present time, in Greece that is a highly seismogenic country, the preseismic "strange attractor like" precursor is determined on a daily basis for a wave length of the Earth's oscillating electric field of **T = 1 day**. The compiled maps are presented regularly in the site www.earthquakeprediction.gr. It is hoped that, soon, two more "strange attractors" will be presented on a daily basis. The first will be the one calculated for the wave length of **T = 14 days** and the second will be the one with the wave length of **T = 6 months**. Actually, it will be a live experiment that can be followed by all scientists being interested in this topic.

It must be pointed out that this methodology is a clearly "time prediction" one while the extent, figure and location of the observed ellipses before any large **EQ** have not yet been related to the actual epicentre location of it. A possible solution towards the epicentre calculation could be worked out using traditional inversion schemes as follows:

Any sample point of an ellipsis is calculated by the intersection of two vectors of the oscillating electric field observed at a time (t), at two different monitoring sites **M1, M2.** That oscillating electric field is generated by the focal area of a large **EQ** of a specific epicentral area. Therefore, at a specific time **(t)** the coordinates of the epicentre of the large **EQ** can be expressed in terms of the coordinates of the monitoring sites and the coordinates of the ellipsis sample point through their interconnecting functions related to the potential spatial distribution from the focal area. Furthermore, the entire scheme will be constrained so that the electric field vector intersections, sample points, obey the basic function properties of an ellipsis. Then, the entire forward model (Thanassoulas et al. 2008a) can be inverted by any suitable mathematical routine (**SVD**) in order to determine the best **EQ** epicentre parameters (**x, y** in a Cartesian system) in a least squares sense (**LSQ**) for all recorded sample points. We hope that a young researcher will have the opportunity and will to try to implement it in the near future.

Finally the specific methodology of the "strange attractor like" precursor will be applied and tested on longer wave lengths (**T > 6 months**) as soon as the required data will be available from **PYR** and **HIO** monitoring sites.

## 5. References.